\documentclass[conference,final]{IEEEtran}
\usepackage{amsmath,amssymb}
\usepackage{graphicx}
\newcommand{\LB}{\left(}
\newcommand{\RB}{\right)}
\newcommand{\LSB}{\left[}
\newcommand{\RSB}{\right]}
\newcommand{\htp}{^{\sf H}}
\newcommand{\tp}{^{\sf T}}

\newfont{\bbb}{msbm10 scaled 500}

\newfont{\bb}{msbm10 scaled 1100}
\newcommand{\CC}{\mbox{\bb C}}

\newcommand{\hv}{{\bf h}}

\newcommand{\nv}{{\bf n}}

\newcommand{\rv}{{\bf r}}

\newcommand{\uv}{{\bf u}}
\newcommand{\wv}{{\bf w}}
\newcommand{\vv}{{\bf v}}
\newcommand{\xv}{{\bf x}}
\newcommand{\yv}{{\bf y}}

\newcommand{\zerov}{{\bf 0}}

\newcommand{\Am}{{\bf A}}

\newcommand{\Dm}{{\bf D}}

\newcommand{\Hm}{{\bf H}}
\newcommand{\Id}{{\bf I}}
\newcommand{\Jm}{{\bf J}}

\newcommand{\Qm}{{\bf Q}}
\newcommand{\Rm}{{\bf R}}
\newcommand{\Sm}{{\bf S}}
\newcommand{\Tm}{{\bf T}}

\newcommand{\Wm}{{\bf W}}

\newcommand{\Zm}{{\bf Z}}

\newcommand{\Cc}{{\cal C}}

\newcommand{\Nc}{{\cal N}}

\newcommand{\deltav}{\hbox{\boldmath$\delta$}}

\newcommand{\Phim}{\hbox{\boldmath$\Phi$}}

\newcommand{\Thetam}{\hbox{\boldmath$\Theta$}}

\newcommand{\trace}{{\hbox{tr}\,}}

\renewcommand{\Re}{{\rm Re}}

\newcommand{\defines}{{\,\,\stackrel{\scriptscriptstyle \bigtriangleup}{=}\,\,}}

\newtheorem{theorem}{Theorem}%[section]
\newtheorem{cor}{Corollary}

\newtheorem{remark}{Remark}[section] 

\newtheorem{assumption}{\indent \bf A}

\begin{document}
\title{Massive MIMO: How many antennas do we need?}
\author{
\IEEEauthorblockN{Jakob Hoydis\IEEEauthorrefmark{1}\IEEEauthorrefmark{2}, Stephan ten Brink\IEEEauthorrefmark{3}, and M\'{e}rouane Debbah\IEEEauthorrefmark{2}} 
\IEEEauthorblockA{\IEEEauthorrefmark{1}Department of Telecommunications, Sup\'{e}lec, 3 rue Joliot-Curie, 91192 Gif-sur-Yvette, France}
\IEEEauthorblockA{\IEEEauthorrefmark{2}Alcatel-Lucent Chair on Flexible Radio, Sup\'{e}lec, 3 rue Joliot-Curie, 91192 Gif-sur-Yvette, France}
\IEEEauthorblockA{\IEEEauthorrefmark{3}Bell Labs, Alcatel-Lucent, Lorenzstr. 10, 70435 Stuttgart, Germany\\
\{jakob.hoydis, merouane.debbah\}@supelec.fr,\ stephan.tenbrink@alcatel-lucent.com}
}
\maketitle

\begin{abstract}
We consider a multicell MIMO uplink channel where each base station (BS) is equipped with a large number of antennas $N$. The BSs are assumed to estimate their channels based on pilot sequences sent by the user terminals (UTs). Recent work has shown that, as $N\to\infty$, (i) the simplest form of user detection, i.e., the matched filter (MF), becomes optimal, (ii) the transmit power per UT can be made arbitrarily small, (iii) the system performance is limited by pilot contamination. The aim of this paper is to assess to which extent the above conclusions hold true for large, but finite $N$. In particular, we derive how many antennas per UT are needed to achieve $\eta\,$\% of the ultimate performance. We then study how much can be gained through more sophisticated minimum-mean-square-error (MMSE) detection and how many more antennas are needed with the MF to achieve the same performance. Our analysis relies on novel results from random matrix theory which allow us to derive tight approximations of achievable rates with a class of linear receivers.
\end{abstract}

\section{Introduction}
As wireless networks are inherently limited by their own interference, a lot of research focuses on interference reduction techniques, such as multiuser MIMO \cite{gesbert07}, interference alignment \cite{ali08} or multicell processing \cite{gesbert2010}. Although these techniques can lead to considerable performance gains, it is unlikely that they will be able to carry the exponentially growing wireless data traffic. Due to this reason, a significant network densification, i.e., increasing the number of antennas per unit area, seems inevitable. One way of densifying the network consists in cell-size shrinking, such as the deployment of femto or small cells  \cite{hoydis11}, which comes at the cost of additional equipment and increased interference. Another, seemingly simpler, but also less explored option is the use of very large antennas arrays at the base stations (BSs) \cite{marzetta10}. However, it is well known that accurate channel state information (CSI) must be acquired to reap the benefits of additional antennas. This poses, in particular in fast fading channels, a challenge as the number of antennas grows \cite{caire10}. Thus, \emph{massive MIMO} is only feasible in time-division duplex (TDD) systems where channel reciprocity can be exploited. Here, the channels are estimated based on orthogonal pilot tones which are sent by the user terminals (UTs). In such systems, additional antennas do not increase the training overhead and, therefore, ``always help'' \cite{marzetta06}. Nevertheless, for a given coherence time, the number of orthogonal pilot sequences is limited and they must be reused in adjacent cells. This leads to channel estimation impairments known as ``pilot contamination'' \cite{jose11b}.

The use of massively many antennas was advocated for the first time in \cite{marzetta10} and has since then received growing research interest \cite{jose11a,ngo2011conf,gopalakrishnan11}. It is particularly intriguing that with an infinite number of antennas, the simplest forms of user detection and precoding, i.e., matched filtering (MF) and eigenbeamforming, become optimal, the transmit power can be made arbitrarily small, and the performance is ultimately limited by pilot contamination. Therefore, several works address the problem of how the negative impact of pilot contamination can be reduced \cite{jose11a,jose11b}.

But what is the difference between massive MIMO and classical MIMO techniques whose benefits are well-known since more than a decade \cite{telatar1999}? This paper tries to give an answer to this question. We first provide a definition of massive MIMO as an operating condition of cellular systems where multiuser interference and noise are small compared to pilot contamination. Whether or not this condition is satisfied depends in general on several factors: the number of BS antennas $N$, the number of UTs per degree of freedom (DoF) offered by the channel (we denote by DoF the rank of the antenna correlation matrix which might be smaller than $N$), the signal-to-noise ratio (SNR) and the path loss. We then study if more antennas can compensate for suboptimal user detection, i.e., how many more antennas does the MF need to achieve minimum-mean-square-error (MMSE) detection performance. Interestingly, both detectors achieve the same performance with an unlimited number of BS antennas.

The technical contributions of this work are summarized as follows: We consider a very general channel model which allows us to assign a different path loss and  receive antenna correlation matrix to each channel from a UT to a BS. Assuming a large system limit where the number of BS antennas $N$ and the number of UTs per cell $K$ grow infinitely large at the same speed, we derive asymptotically tight approximations of ergodic achievable rates with a class of linear receivers, accounting explicitly for channel training and pilot contamination. Simulations suggest that our approximations are valid for even small $N$ and $K$. Our analysis is different from \cite{marzetta10} which assumes that $K/N\to 0$. However, we obtain their results as a special case. Although we do not fully exploit the generality of the channel model in the current paper, it enables future studies on massive MIMO under more realistic channel models with the goal of analyzing  the impact of antenna correlation, spacing and aperture. Lastly, the techniques developed in this work can be directly applied to the downlink \cite{hoydis11b}.

\section{System model}
Consider a multi-cellular system consisting of $L>1$ cells with one BS and $K$ UTs in each cell. Each BSs is equipped with $N$ antennas, the UTs have a single antenna. The focus of this paper is on the uplink without any form of BS cooperation. The received baseband signal vector $\yv_j\in\CC^{N}$ at BS~$j$ at a given time reads
\begin{align}
 \yv_j = \sqrt{\rho}\sum_{l=1}^L\Hm_{jl}\xv_l + \nv_j
\end{align}
where $\Hm_{jl}\in\CC^{N\times K}$ is the channel matrix from the UTs in cell $l$ to BS~$j$, $\xv_l=[x_{l1}\cdots x_{lK}]\tp\sim\Cc\Nc\LB\zerov,\Id_K\RB$ is the message-bearing transmit vector from the  UTs in cell $l$, $\nv_j\sim\Cc\Nc\LB\zerov,\Id_N\RB$ is a noise vector and $\rho>0$ denotes the transmit SNR. We model the $k$th column vector $\hv_{jlk}\in\CC^{N}$ of the matrix $\Hm_{jl}$ as 
\begin{align}\label{eq:channelmodel}
 \hv_{jlk} = \tilde{\Rm}_{jlk}\wv_{jlk}
\end{align}
where $\Rm_{jlk}\defines\tilde{\Rm}_{jlk}\tilde{\Rm}_{jlk}\htp\in\CC^{N\times N}$ are deterministic correlation matrices and $\wv_{jlk}\sim\Cc\Nc\LB\zerov,\Id_N\RB$ are fast fading channel vectors. The above channel model is very versatile as it allows us to assign a different antenna correlation (including path loss) to each channel vector. Moreover, the matrices $\Rm_{jlk}$ do not need to have full rank. This is especially important for large antenna arrays with a significant amount of antenna correlation due to either insufficient antenna spacing or a lack of scattering. 

\vspace{5pt}
\begin{remark}\label{rem:channelmodel}
In particular, \eqref{eq:channelmodel} can be used to represent a physical channel model with a fixed number of dimensions or angular bins $P$ as in \cite{ngo2011conf}, by letting $ \tilde{\Rm}_{jlk} = \sqrt{\ell_{jlk}}\LSB\Am\ \zerov_{N\times N-P}\RSB $,  
where $\Am\in\CC^{N\times P}$ and $\ell_{jlk}$ is the inverse path loss from UT $k$ in cell $l$ to BS $j$. We will use a particular form of this model in Section~\ref{sec:massiveMIMO}.
\end{remark}

\subsection{Channel estimation}
During a dedicated training phase (whose length we ignore in this work), the UTs in each cell transmit orthogonal pilot sequences which allow the BSs to compute estimates $\hat{\Hm}_{jj}=[\hat{\hv}_{jj1}\cdots,\hat{\hv}_{jjK}]\in\CC^{N\times K}$ of their local channels $\Hm_{jj}$. The same set of orthogonal pilot sequences is reused in every cell so that the channel estimate is corrupted by pilot contamination from adjacent cells \cite{jose09,marzetta10}.  Under these assumptions, BS $j$ estimates the channel vector $\hv_{jjk}$ based on the observation
\begin{align}\label{eq:chnest}
 \yv^\tau_{jk} = \hv_{jjk} + \sum_{l\ne j} \hv_{jlk} + \frac1{\sqrt{\rho_\tau}}\nv_{jk}
\end{align}
where $\nv_{jk}\sim\Cc\Nc\LB\zerov,\Id_N\RB$ and $\rho_\tau>0$ is the effective training SNR. Assuming MMSE estimation \cite{verdu_book}, we can decompose the channel as $\hv_{jjk}=\hat{\hv}_{jjk} + \tilde{\hv}_{jjk}$, where $\hat{\hv}_{jjk}\sim\Cc\Nc\LB\zerov,\Phim_{jjk}\RB$ is the channel estimate, $\tilde{\hv}_{jjk}\sim\Cc\Nc\LB\zerov,\Rm_{jjk}-\Phim_{jjk}\RB$ is the independent estimation error and the matrices $\Phim_{jlk}$ are defined as $(1\le j,l \le L, 1\le k \le K)$:
\begin{align}
 \Phim_{jlk} &= \Rm_{jjk}\Qm_{jk}\Rm_{jlk}\\
 \Qm_{jk} &= \LB\frac{1}{\rho_\tau}\Id_N + \sum_{l}\Rm_{jlk}\RB^{-1}.
\end{align}

\subsection{Achievable rates with linear detection}
We consider linear single-user detection, where the $j$th BS estimates the symbols $x_{jm}$, $m=1,\dots,K$, of the UTs in its cell by calculating the inner products between the received vector $\yv_j$ and the linear filters $\rv_{jm}\in\CC^N$. Two particular detectors are of practical interest, the matched filter $\rv^\text{MF}_{jm}$ and the MMSE detector $\rv_{jm}^\text{MMSE}$ which we define as
\begin{align}
 \rv^\text{MF}_{jm} & = \hat{\hv}_{jjm}\\
\rv^\text{MMSE}_{jm} & = \LB\hat{\Hm}_{jj}\hat{\Hm}_{jj}\htp + \Zm_j + N\lambda\Id_N\RB^{-1}\hat{\hv}_{jjm}
\end{align}
where $\lambda>0$ is a design parameter and
\begin{align}\nonumber
 \Zm_j&=\mathbb{E}\LSB\tilde{\Hm}_{jj}\tilde{\Hm}_{jj}\htp + \sum_{l\ne j}\Hm_{jl}\Hm_{jl}\RSB \\
&=\sum_{k}\LB\Rm_{jjk}-\Phim_{jjk}\RB+\sum_{l\ne j}\sum_{k} \Rm_{jlk}.
\end{align}

\vspace{5pt}
\begin{remark}
Note that a BS could theoretically estimate all channel matrices $\Hm_{jl}$ from the observations \eqref{eq:chnest} to further improve the performance. Nevertheless, high path loss to neighboring cells is likely to render these channel estimates unreliable and the potential performance gains are expected to be marginal. Our formulation of $\rv^\text{MMSE}_{jm}$ further allows us to treat $\lambda$ (and also $\Zm_j$) as a design parameter which could be optimized. A natural choice is $\lambda=\frac{1}{\rho N}$.
\end{remark}
\vspace{5pt}

Using a standard bound based on worst-case uncorrelated additive noise  yields the ergodic achievable rate $R_{jm}$ of UT $m$ in cell $j$ \cite{hassibi03}:
\begin{align}
 R_{jm} = \mathbb{E}\LSB\log_2\LB1+\gamma_{jm}\RB\RSB
\end{align}
where the associated ``signal-to-interference-plus-noise ratio'' (SINR) $\gamma_{jm}$ is given by \eqref{eq:sinr} on the top of the next page.

\begin{figure*}
 \begin{align}\label{eq:sinr}
 \gamma_{jm} = \frac{\left|\rv_{jm}\htp\hat{\hv}_{jjm}\right|^2}{\mathbb{E}\LSB\rv_{jm}\htp\LB\frac{1}{\rho}\Id_N + \tilde{\hv}_{jjm}\tilde{\hv}_{jjm}\htp -\hv_{jjm}\hv_{jjm}\htp  +\sum_{l} \Hm_{jl}\Hm_{jl}\htp \RB\rv_{jm}\,|\hat{\Hm}_{jj}\RSB}
\end{align}
\begin{center}
\line(1,0){180}
\end{center}
\end{figure*}
 
\section{Asymptotic analysis}\label{sec:asym}
In this section, we present our main technical results. All proofs are provided in \cite{hoydis11b}. Under the assumption that $N$ and $K$ grow infinitely large while keeping a finite ratio $K/N$, we will derive deterministic approximations $\bar{\gamma}_{jm}$ of $\gamma_{jm}$, such that
\begin{align}\label{eq:conv_sinr}
 \gamma_{jm} - \bar{\gamma}_{jm} \xrightarrow[]{\text{a.s.}} 0
\end{align}
where `$ \xrightarrow[]{\text{a.s.}}$' denotes almost sure convergence. We will refer to the quantities $\bar{\gamma}_{jm}$ as \emph{deterministic equivalents} of $\gamma_{jm}$. The large system analysis implicitly requires that the channel coherence times scales linearly with $K$ (to allow for a sufficiently large number of orthogonal pilot sequences). However, since we will use our results as approximations for realistic system dimensions, this assumption does not pose any problem.

We would further like to remark that, in the asymptotic limit, the term inside the expectation of the denominator of $\gamma_{jm}$ \eqref{eq:sinr} can be arbitrarily closely approximated by a deterministic quantity and we do not need to compute the expectation explicitly. Moreover, since $\mathbb{E}[|\gamma_{jm}|]$ and $\bar{\gamma}_{jm}$ are uniformly bounded, the almost sure convergence in \eqref{eq:conv_sinr} implies by \cite{billingsley}[Corollary, p. 218]:
\begin{align}
 R_{jm} - \log_2\LB1+\bar{\gamma}_{jm}\RB\xrightarrow[]{} 0.
\end{align}

In the sequel, we assume that the following technical conditions hold:

\vspace{5pt}
\begin{assumption}
$\limsup_N \lVert\Rm_{jlk}\rVert<\infty$ for all $j,l,k$.
\end{assumption}
\vspace{5pt}

\begin{assumption}
$\liminf_N \frac1N\trace\Rm_{j,l,k} > 0$ for all $j,l,k$.
\end{assumption}
\vspace{5pt}

Our first result is a deterministic equivalent of the SINR at the output of the matched filter:

\vspace{5pt}
\begin{theorem}[Matched filter]\label{th:mf}
A deterministic equivalent of $\gamma_{jm}$ for the matched filter is given as
 \begin{align*}
 &\bar{\gamma}^\text{MF}_{jm} =\\
& \frac{\LB\frac1N\trace \Phim_{jjm}\RB^2}{\frac1{\rho N^2}\trace\Phim_{jjm}+\frac1N\sum_{l,k} \frac1N\trace\Rm_{jlk}\Phim_{jjm} + \sum_{l\ne j}\left|\frac1N\trace\Phim_{jlm}\right|^2}.
\end{align*}
\end{theorem}
\vspace{5pt}

Our second result is a deterministic equivalent of the SINR with MMSE detection:

\vspace{5pt}
\begin{theorem}[MMSE detector]\label{th:mmse}
A deterministic equivalent of $\gamma_{jm}$ for the MMSE detector is given as
\begin{align*}
\bar{\gamma}_{jm}^\text{MMSE} = \frac{\delta_{jm}^2}{\frac1{\rho N^2}\trace\Phim_{jjm}\bar{\Tm}_j'+ \frac1N\sum_{l,k}\mu_{jlkm} + \sum_{l\ne j}\left|\vartheta_{jlm}\right|^2}
\end{align*}
where
\begin{align*}
\vartheta_{jlk} &= \frac1N\trace\Phim_{jlk}\Tm_j\\
\vartheta_{jlkm}' &=\frac1N\trace\Phim_{jlk}\Tm_{jm}'\\
\mu_{jlkm} &=  \frac1N\trace\Rm_{jlk}{\Tm}'_{jm}\\
&\quad-\frac{2\Re\LB\vartheta_{jlk}^*\vartheta_{jlkm}'\RB\LB1+\delta_{jk}\RB - \left|\vartheta_{jlk}\right|^2\delta_{jkm}'}{\LB1+\delta_{jk}\RB^2}
\end{align*}
and where
\begin{itemize}
 \item $\Tm_j=\Tm(\lambda)$ and $\deltav_{j}=[\delta_{j1}\cdots\delta_{jK}]\tp=\deltav(\lambda)$ are given by Theorem~\ref{th:detequ} for $\Dm=\Id_N$, $\Sm=\Zm_j/N$, $\Rm_k=\Phim_{jjk}\, \forall k$,
\item  $\bar{\Tm}_j'=\Tm'(\lambda)$ is given by Theorem~\ref{th:detequder} for $\Dm=\Id_N$, $\Sm=\Zm_j/N$, $\Thetam=\Id_N$, $\Rm_k=\Phim_{jjk}\, \forall k$,
\item ${\Tm}'_{jm}=\Tm'(\lambda)$, ${\deltav}_{jm}'=\LSB\delta_{j1m}'\cdots\delta_{jKm}'\RSB\tp = \deltav'(\lambda)$ are given by Theorem~\ref{th:detequder} for $\Dm=\Id_N$, $\Sm=\Zm_j/N$,$\Thetam=\Phim_{jjm}$, $\Rm_k=\Phim_{jjk}\, \forall k$.
\end{itemize}
Theorems~\ref{th:detequ} and \ref{th:detequder} can be found in the Appendix.
\end{theorem}
\vspace{5pt}

\begin{remark}
 The theorem can be drastically simplified if a less general channel model is considered, e.g. the same correlation matrices $\Rm_{jlk}$ for all UTs, no correlation and only path loss, $\Zm_j=\zerov$. Due to space reasons, we only state the most general form here and provide one special case (cf. Corollary \ref{cor:mmse}) where $\bar{\gamma}^\text{MMSE}_{jm}$ can be even given closed form.
\end{remark}
\vspace{5pt}

Interestingly, the performances of matched filter and MMSE detector coincide for an infinite number of antennas:

\vspace{5pt}
\begin{cor}[Performance with infinitely many antennas]\label{cor:limit}
\begin{align*}
   \bar{\gamma}^\text{MF}_{jm},\ \bar{\gamma}^\text{MMSE}_{jm} \xrightarrow[N\to\infty,\ K/N\to 0]{} \bar{\gamma}^\infty_{jm}\defines \frac{\beta_{jjm}^2}{\sum_{l\ne j}|\beta_{jlm}|^2}
\end{align*}
where $\beta_{jlk}=\lim_N\frac1N\trace\Phim_{jlk}$, whenever the limit exists.
\end{cor}
\vspace{5pt}

Note that $\bar{\gamma}^\infty_{jm}$ corresponds also to the asymptotic signal-to-interference ratio (SIR) derived in \cite[Eq. (13)]{marzetta10}.

\section{On the massive MIMO effect}\label{sec:massiveMIMO}
Let us for now ignore the effects of estimation noise, i.e., $\rho_\tau\to\infty$, and consider the simple channel model
\begin{align}\label{eq:chnmodelsimple}
 \Hm_{jj} = \sqrt{\frac{N}{P}}\Am\Wm_{jj},\quad  \Hm_{jl} = \sqrt{\alpha\frac{N}{P}}\Am\Wm_{jl},\quad l\ne j
\end{align}
where $\Am\in\CC^{N\times P}$ is composed of $P\le N$ columns of an arbitrary unitary $N\times N$ matrix, $\Wm_{jl}\in\CC^{P\times K}$ are standard complex Gaussian matrices and $\alpha\in(0,1]$ is an intercell interference factor. Note that this is a special case of \eqref{eq:channelmodel}. Under this model, the total energy of the channel grows linearly with the number of antennas $N$ and UTs $K$, since
$ \mathbb{E}\LSB\trace\Hm_{jj}\Hm_{jj}\htp\RSB = \frac {KN}{P} \trace\Am\Am\htp = KN$.
The motivation behind this channel model is twofold. First, we assume that the antenna aperture increases with each additional antenna element. Thus, the captured energy increases linearly with $N$. This is in contrast to existing works which assume that more and more antenna elements are packed into a fixed volume, see e.g. \cite{gesbert02,wei05}. An insufficiency of this channel model is that the captured energy grows without bounds as $N\to\infty$. However, we believe that linear energy gains can be achieved up to very large numbers of antennas if the size of the antenna array is scaled accordingly. Second, the number of DoF $P$ offered by the channel does not need to be equal to $N$ \cite{ngo2011conf}. One could either assume $P$ to be large but constant or to scale with $N$, e.g. $P=cN$, where $c\in(0,1]$. In general, $P$ depends on the amount of scattering in the channel and, therefore, on the radio environment. Under the assumptions made above, we obtain the following corollary from Theorem~\ref{th:mf}:

\vspace{5pt}
\begin{cor}\label{cor:mf}
  For the channel model in \eqref{eq:chnmodelsimple} and $\rho_\tau\to\infty$, $\bar{\gamma}_{km}^\text{MF}$, for all $k,m$, can be given in closed form as 
\begin{align}\label{eq:asymsinr}
 \bar{\gamma}^\text{MF}\ \defines\ \frac{1}{\underbrace{\frac{\bar{L}}{\rho N}}_{\text{noise}}\ + \underbrace{\frac KP \bar{L}^2}_{\text{multiuser interference}} + \underbrace{\alpha(\bar{L}-1)}_{\text{pilot contamination}}}
\end{align}
where $\bar{L}=1+\alpha(L-1)$.
\end{cor}
\vspace{5pt}

We define the associated rate $\bar{R}^\text{MF}$ as
\begin{align}\label{eq:mfrate}
 \bar{R}^\text{MF}=\log_2\LB1+\bar{\gamma}^\text{MF}\RB.
\end{align}

One can make several observations from \eqref{eq:asymsinr}. Obviously, the effective SNR $\rho N$ increases linearly with $N$. Thus, if the number of antennas is doubled, the transmit power can be reduced by a factor of two to achieve the same SNR. Less obvious is that the multiuser interference depends mainly on the ratio $P/K$ (number of DoF per UT) and not directly on the number of antennas. Moreover, noise and multiuser interference vanish for $N,P\to\infty$ while pilot contamination is the only performance-limiting factor \cite{marzetta10}:
\begin{align}
 \bar{\gamma}^\text{MF}\xrightarrow[N,P\to\infty, \ K/N\to 0]{}\gamma_\infty =\frac{1}{\alpha(\bar{L}-1)}.
\end{align}
We denote by $R_\infty$ the ultimately achievable rate, defined as
\begin{align}
 R_\infty = \log_2\LB1+\gamma_\infty\RB=\log_2\LB1+\frac{1}{\alpha(\bar{L}-1)}\RB.
\end{align}
It is interesting that even with more sophisticated linear single-user detection, such as MMSE detection, the ultimate performance limit $\gamma_\infty$ cannot be exceeded  (see Corollary~\ref{cor:limit} and \cite{ngo2011conf}). Note that without pilot contamination, i.e., for $L=1$ or $\alpha=0$, the SINR grows without bounds as $P,N\to\infty$. If $P$ is fixed but large, the SINR saturates at a smaller value than $\gamma_\infty$. In this case, adding additional antennas only improves the SNR but does not reduce the multiuser interference. Thus, with a finite number of DoF, MMSE detection remains also for $N\to\infty$ superior to MF.

Based on the above observations, we believe that it is justified to speak about a \emph{massive MIMO effect} whenever $\gamma_{jm}$ is close to $\gamma_\infty$, or in other words, whenever noise and multiuser interference are small compared to the pilot contamination. It becomes evident from \eqref{eq:asymsinr} that the number of antennas needed for massive MIMO depends strongly on the system parameters $P$, $K$, $L$, $\alpha$ and $\rho$. In particular, there is no massive MIMO effect without pilot contamination since $\gamma_\infty\to\infty$. Thus, massive MIMO can be seen as a particular operating condition in multi-cellular systems where the performance is ultimately limited by pilot contamination and the matched filter achieves a performance close to this ultimate limit. To make this definition more precise, we say that we operate under massive MIMO conditions if, for some desired $\eta\in(0,1)$,
\begin{align}\label{eq:massiveMIMOcond}
 \bar{R}^\text{MF}\ \ge\ \eta R_\infty.
\end{align}
 This condition implies that we achieve at least the fraction $\eta$ of the ultimate performance with the simplest form of single-user detection. Replacing $\bar{R}^\text{MF}$ in the last equation by \eqref{eq:mfrate} and solving for $P/K$ leads to
\begin{align}
 \frac PK \ge \LB\frac1{\bar{L}^2\LSB\LB1+\gamma_\infty\RB^\eta - 1\RSB} - \frac1{\rho N\bar{L}} - \frac{\alpha (\bar{L}-1)}{\bar{L}^2}\RB^{-1}.
\end{align}

Thus for a given set of parameters $\rho$, $N$, $\alpha$ and $L$, we can find the fraction $P/K$ necessary to satisfy \eqref{eq:massiveMIMOcond}.
For the particular channel model considered in this section, also Theorem~\ref{th:mmse} for the MMSE detector can be significantly simplified:

\vspace{5pt}
\begin{cor}\label{cor:mmse}
 For the channel model in \eqref{eq:chnmodelsimple} and $\rho_\tau\to\infty$, $\bar{\gamma}^\text{MMSE}_{km}$, for all $k,m$, can be given in closed-form as
\begin{align*}
 \bar{\gamma}^\text{MMSE} = \frac{1}{\frac{\bar{L}}{\rho N}X + \frac KP \bar{L}^2Y + \alpha\LB\bar{L}-1\RB}
\end{align*}
where $\bar{L}=1 + \alpha(L-1)$ and
\begin{align*}
X &= \frac{Z^2}{Z^2-\frac KP}\\
Y &= X + \frac{\LSB1+\alpha^2(L-1)\RSB(1-2Z)}{\bar{L}^2\LB Z^2-\frac KP\RB}\\
Z &= \lambda\bar{L}(1+\delta) + \frac KP\LSB1+(1+\delta)\LB\bar{L}^2-1\RB\RSB\\
\delta &= \frac{1-\lambda\bar{L} - \frac KP \bar{L}^2 + \sqrt{\LB1+\lambda\bar{L} + \frac KP \bar{L}^2\RB^2 - 4\frac KP}}{2\LSB\lambda\bar{L} + \frac KP \LB\bar{L}^2-1\RB\RSB}.
\end{align*} 
\end{cor}
\vspace{5pt}

The last result reveals that also the SINR of the MMSE detector depends on $P,K,N$ and $\rho$ only through the ratio $P/K$ and the effective SNR $\rho N$. Hence, we can use Corollary~\ref{cor:mmse} to determine the ratio $P/K$ necessary to satisfy the condition 
\begin{align}\label{eq:condMMSE}
 \bar{R}^\text{MMSE}=\log\LB1+\bar{\gamma}^\text{MMSE}\RB\ge\eta R_\infty
\end{align}
which can be very efficiently solved by a simple line search, e.g. via bisection. We assume in the sequel $\lambda=1/(\rho N)$.

\begin{figure}
\centering
\includegraphics[width=0.485\textwidth]{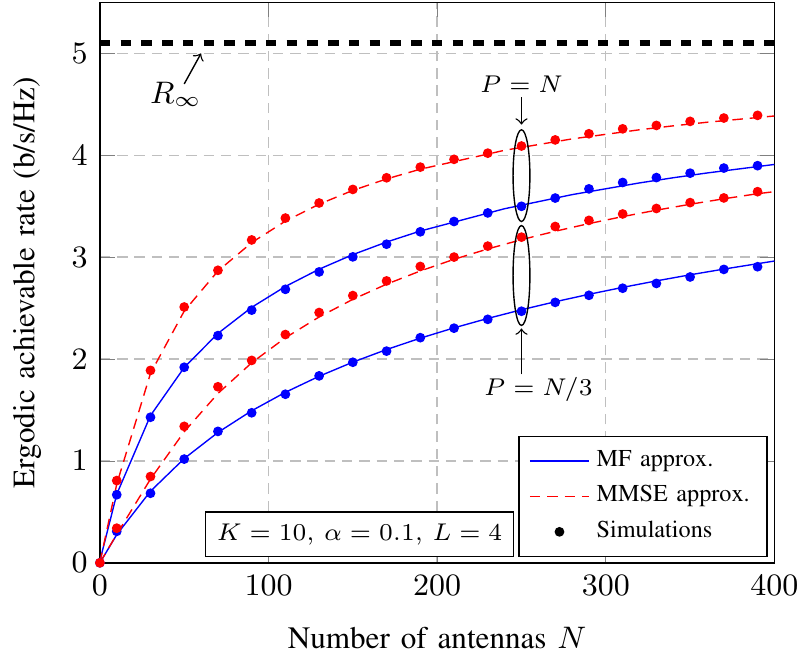}
\caption{Ergodic achievable rate with MF and MMSE detection versus number of antennas $N$ for $P=\{N,N/3\}$ for $\rho=0\, \text{dB}$.\label{fig:rate_vs_N}}
\end{figure}

Before we proceed, let us verify the accuracy of the approximations $\bar{R}^\text{MF}$ and $\bar{R}^\text{MMSE}$ for finite $(N,K)$. In Fig.~\ref{fig:rate_vs_N}, we depict the ergodic achievable rate $R_{jm}$ of an arbitrary UT with MF and MMSE detection as a function of the number of antennas $N$ for $K=10$ UTs, $L=4$ cells, $\rho=0\,\text{dB}$ and intercell interference factor $\alpha=0.1$. We compare two different cases: $P=N$ and $P=N/3$. As expected, the performance in the latter scenario is inferior due to stronger multiuser interference. Most importantly, our closed-form approximations are almost indistinguishable from the simulation results over the entire range of $N$.

\begin{figure}
\centering
\includegraphics[width=0.485\textwidth]{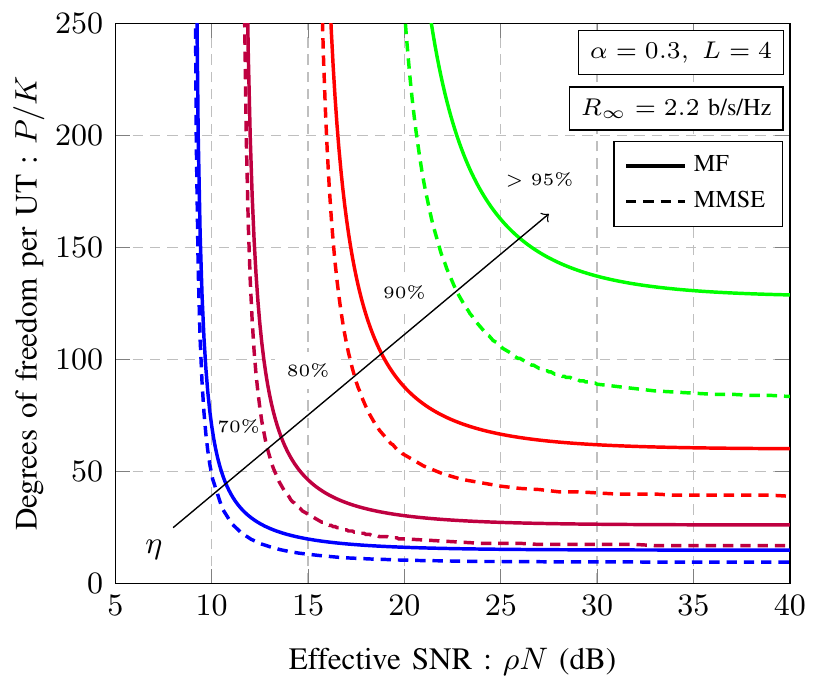}
\caption{Degrees of freedom per UT $P/K$ necessary to achieve $\eta R_\infty$ versus effective SNR $\rho N$ for $L=4$ and $\alpha=0.3$.\label{fig:massiveMIMO1}}
\end{figure}

\begin{figure}
\centering
  \includegraphics[width=0.485\textwidth]{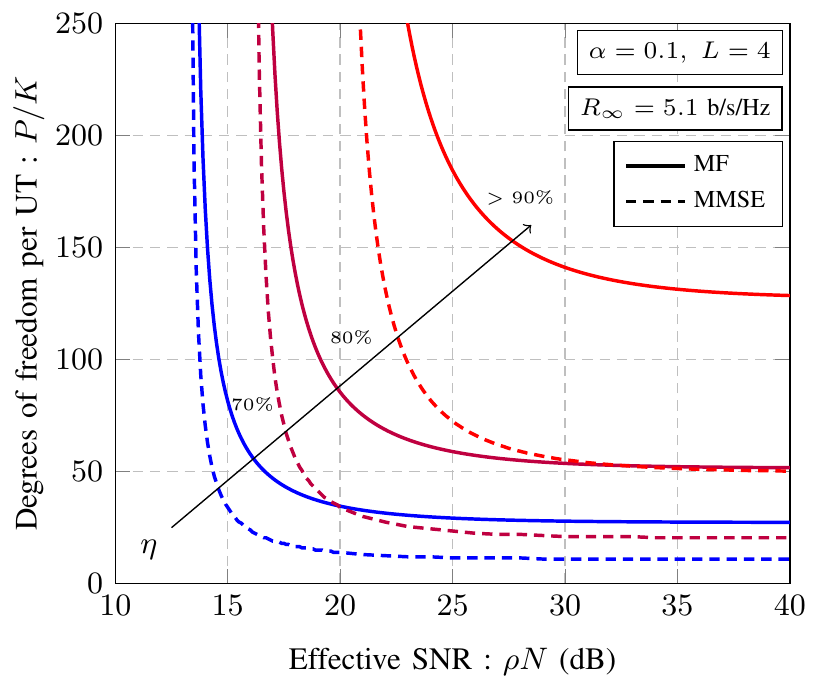}
\caption{Degrees of freedom per UT $P/K$ necessary to achieve $\eta R_\infty$ versus effective SNR $\rho N$ for $L=4$ and $\alpha=0.1$.\label{fig:massiveMIMO2}}
\end{figure}

Figs.~\ref{fig:massiveMIMO1} and \ref{fig:massiveMIMO2} show the necessary DoF per UT $P/K$ for a given effective SNR $\rho N$ to achieve a spectral efficiency of $\eta R_\infty$ with either the MF (solid lines) or MMSE detector (dashed lines). We consider $L=4$ cells and an intercell interference factor $\alpha=0.3$ and $\alpha=0.1$, respectively. The plots must be understood in the following way: Each curve corresponds to a particular value of $\eta$. In the region above each curve, the condition \eqref{eq:massiveMIMOcond}, respectively \eqref{eq:condMMSE}, is satisfied.

Let us first focus on Fig.~\ref{fig:massiveMIMO1} with $\alpha=0.3$. For an effective SNR $\rho N = 20\, \text{dB}$ (e.g. $\rho=0\,\text{dB}$ and $N=100=20\, \text{dB}$), we need about $P/K=90$ DoF per UT to achieve $90\,$\% of the ultimate performance $R_\infty$, i.e., $0.9\times 2.2 \approx 2\,\text{b/s/Hz}$. If $P\approx N$, only a single UT could be served (Note that this is a simplifying example. Our analysis assumes $K\gg 1$.). However, if we had $N=1000=30\,\text{dB}$ antennas, the transmit power $\rho$ could be decreased by $10\, $dB and $10$ UTs could be served with the same performance. At the same operating point, the MMSE detector requires only $\sim 60$ DoF per UT to achieve $90\,$\% of the ultimate performance. Thus, MMSE detection would allow us to increase the number of simultaneously served UTs by a factor $\frac{90}{60}=1.5$.
This example also demonstrates the importance of the relation between $N$ and $P$. In particular, if $P$ saturates for some $N$, adding more antennas increases the effective SNR but does not reduce the multiuser interference. Thus, the number of UTs which can be simultaneously supported depends significantly on the radio environment. We can further see that adding antennas shows diminishing returns. This is because the distances between the curves for different values of $\eta$ grow exponentially fast. Remember that for $\eta=1$, a ratio of $P/K=\infty$ would be needed. A last observation we can make is that the absolute difference between MF and MMSE detection is marginal for small values of $\eta$ but gets quickly pronounced as $\eta\to 1$.

Turning to Fig.~\ref{fig:massiveMIMO2} for $\alpha=0.1$, we can see that for the same effective SNR $\rho N = 20\, \text{dB}$ and the same number of DoF per UT $P/K=90$ as in the previous example, only $80\,$\% of the ultimate performance are achieved by the MF. However, since the intercell interference is significantly smaller compared to the previous example, this corresponds to $0.9 \times 5.1\approx 4.6\,\text{b/s/Hz}$. Thus, although we operate further away from the ultimate performance limit, the resulting spectral efficiency is still higher. With MMSE detection, only $35$ DoF per UT are necessary to achieve the same performance and, consequently, $90/35\approx 2.5$ times more UTs could be simultaneously served. With decreasing intercell interference (and hence decreasing pilot contamination) the advantages of MMSE detection become more and more important.

\section{Conclusions}
We have studied the uplink of a MIMO multi-cell system where channel estimates are prone to pilot contamination. Under a very general channel model, allowing for an individual correlation matrix for each UT, and the assumption that the number of BS antennas and UTs grows large, we have derived deterministic equivalents of  achievable rates with matched filter and MMSE detector. These results have been used to study the performance of
both detectors in the large but finite $N$ regime. Interestingly, their performance depends mainly on the DoF per UT the channel offers and the effective SNR. Moreover, we have determined (i) how many antennas are needed to achieve $\eta\,$\% of the ultimate performance limit and (ii) how many more antennas are needed by the MF to achieve MMSE performance. Our results can be also applied for the study of the downlink \cite{hoydis11b} and could be further used to analyze the effects of more realistic channels models, such as antenna correlation, spacing and aperture.

\appendix
\begin{theorem}[{\cite[Theorem 1]{wagner2011}}]\label{th:detequ}
 Let $\Dm\in\CC^{N\times N}$ and $\Sm\in\CC^{N\times N}$ be Hermitian nonnegative definite matrices and let $\Hm\in\CC^{N\times K}$ be a random matrix with columns $\hv_k=\frac1{\sqrt{N}}\Rm_k^{\frac12}\uv_k$, where $\uv_k\in\CC^{N}$ are random vectors of i.i.d. elements with zero mean, unit variance and finite $8$th order moment, and $\Rm_k\in\CC^{N\times N}$ are deterministic covariance matrices. Assume that $\Dm$, $\Sm$ and $\Rm_k$, $k=1,\dots,K$, have uniformly bounded spectral norms (with respect to $N$). Let $N,K\to\infty$, such that $0\le\lim\inf\frac KN \le \lim\sup \frac KN < \infty$. Then, for any $\rho>0$,
\begin{align*}
 \frac1N\trace\Dm\LB\Hm\Hm\htp +\Sm +\rho\Id_N\RB^{-1} - \frac1N\trace\Dm\Tm(\rho) \xrightarrow[]{\text{a.s.}} 0
\end{align*}
where $\Tm(\rho)\in\CC^{N\times N}$ is defined as
$$ \Tm(\rho) = \LB\frac1N\sum_{k=1}^K\frac{\Rm_k}{1+\delta_k(\rho)}  +\Sm + \rho\Id_N\RB^{-1}$$
and the following set of $K$ implicit equations
$$ \delta_k(\rho)  =  \frac1N \trace\Rm_k\Tm(\rho),\quad k=1,\dots,K$$
has a unique solution $\deltav(\rho)=\LSB\delta_1(\rho)\cdots\delta_K(\rho)\RSB\tp\ge 0$. 
\end{theorem}
\vspace{5pt}

\begin{theorem}\label{th:detequder}
 Let $\Thetam\in\CC^{N\times N}$ be a Hermitian nonnegative definite matrix with uniformly bounded spectral norm (with respect to $N$). Under the same conditions as in Theorem~\ref{th:detequ},  
\begin{align*}
  &\frac1N\trace\Dm\LB\Hm\Hm\htp + \Sm+\rho\Id_N\RB^{-1}\Thetam\LB\Hm\Hm\htp  +\Sm+\rho\Id_N\RB^{-1}\\& \qquad- \frac1N\trace\Dm\Tm'(\rho) \xrightarrow[]{\text{a.s.}} 0
\end{align*}
where $\Tm'(\rho)\in\CC^{N\times N}$ is defined as
$$ \Tm'(\rho) = \Tm(\rho)\Thetam \Tm(\rho) + \Tm(\rho)\frac1N\sum_{k=1}^K\frac{\Rm_k\delta'_k(\rho) }{\LB1+\delta_k(\rho)\RB^2}\Tm(\rho)$$
with $\Tm(\rho)$ and $\delta_k(\rho)$ as defined in Theorem~\ref{th:detequ} and $\deltav'(\rho) = \LSB\delta'_1(\rho)\cdots\delta'_K(\rho)\RSB\tp$ given by
\begin{align*}
 \deltav'(\rho) &= \LB\Id_K - \Jm(\rho)\RB^{-1}\vv(\rho)\\
\LSB\Jm(\rho)\RSB_{kl} &= \frac{\frac1N\trace\Rm_k\Tm(\rho)\Rm_l\Tm(\rho)}{N\LB1+\delta_k(\rho)\RB^2}\\
\LSB\vv(\rho)\RSB_k &= \frac1N\trace\Rm_k\Tm(\rho)\Thetam\Tm(\rho)
\end{align*}
where $\Jm(\rho)\in\CC^{K\times K}$ and $\vv(\rho)\in\CC^{K}$.
\end{theorem}

\vfill
\bibliographystyle{IEEEtran}
\bibliography{IEEEabrv,bibliography}

\end{document}